\documentclass[11pt,a4paper,reqno]{amsart}
\pdfoutput=1

\usepackage[T1]{fontenc}
\usepackage[utf8]{inputenc}
\usepackage[a4paper,margin=1in]{geometry}
\usepackage{lmodern}
\usepackage{amssymb}
\usepackage{microtype}
\usepackage{tikz}
\usetikzlibrary{decorations.pathmorphing}
\usepackage[hidelinks]{hyperref}

\usepackage{orcidlink}
\usepackage[nameinlink,noabbrev]{cleveref}
\usepackage{aliascnt}

\newcommand{\Tr}{\operatorname{Tr}}
\newcommand{\T}{\mathsf T}
\newcommand{\ket}[1]{\lvert #1\rangle}
\newcommand{\bra}[1]{\langle #1\rvert}
\newcommand{\ketbra}[2]{\lvert #1\rangle\langle #2\rvert}

\theoremstyle{plain}
\newtheorem{theorem}{Theorem}
\newaliascnt{lemma}{theorem}
\newtheorem{lemma}[lemma]{Lemma}
\aliascntresetthe{lemma}
\crefname{lemma}{lemma}{lemmas}

\theoremstyle{definition}
\newtheorem{strategy}{Strategy}
\crefname{strategy}{strategy}{strategies}

\title{Linear entanglement suffices for $f$-routing}
\author[Kevin Bogner]{Kevin Bogner\,\orcidlink{0009-0003-5167-7177}\\
 \NoCaseChange{{\normalfont\small COSIC, KU Leuven, Belgium}}\\
 \NoCaseChange{{\normalfont\small\texttt{kevin.bogner@kuleuven.be}}}}

\begin{document}

\begin{abstract}
In $f$-routing, Alice receives an $n$-bit string $x$ and an unknown qubit, and Bob receives an $n$-bit string $y$. They share a state prepared in advance and exchange one simultaneous message each. The party selected by the Boolean function $f(x,y)$ must then recover the unknown qubit without further communication. To perform $f$-routing successfully, Alice and Bob need to share entanglement that scales with $n$ for some functions $f$, but the lower and upper bounds for the entanglement are not tight. This paper shows that $2n+2$ shared EPR pairs suffice for every function $f$. The best previous bound that holds for all functions is a shared state of $2^{O(\sqrt{n\log n})}$ qubits. The new strategy moves the cost from the shared state to the messages: Alice sends an exponentially long message of $2^{n+1}+1$ qubits, while Bob sends only classical bits. When the messages are counted towards the resources as well, the previous strategy remains the cheapest known. The result also limits lower bounds that count only the shared state: they cannot exceed a linear number of EPR pairs. The result is formalized in Lean.
\end{abstract}

\maketitle

\section{Introduction}

Position verification~\cite{cgmo} aims to certify where a prover is located using only timed messages. In a routing protocol~\cite{tagging,gardenhose}, two verifiers send the prover two $n$-bit strings $x$ and $y$ together with an unknown qubit, and the prover must return the qubit to the verifier selected by a Boolean function $f(x,y)$. Two colluding attackers stand between the claimed position and the verifiers, one on each side. The attacker who intercepts the qubit must send her message before the other string reaches her, so she does not yet know $f(x,y)$. She cannot keep the qubit and also forward a copy, because an unknown quantum state cannot be copied~\cite{wootterszurek}. Attackers who share enough entanglement in advance can still break every protocol of this kind~\cite{nogo,portbased}. The protocol is therefore secure only against attackers with limited entanglement, and the question is how much an attack needs~\cite{gardenhose}.

The attackers' task is called $f$-routing~\cite{gardenhose,relating}. Alice, the attacker who intercepts the qubit $Q$, receives $x\in\{0,1\}^n$, while Bob receives $y\in\{0,1\}^n$. Beforehand, they share an entangled state. They exchange one simultaneous message each and then recover $Q$ locally, Alice when $f(x,y)=0$ and Bob when $f(x,y)=1$. These two possibilities are the routing cases. A strategy for $f$-routing is a shared state together with the operations that Alice and Bob apply. \Cref{sec:model} defines the task precisely.

How much entanglement $f$-routing needs is known only in part. For upper bounds, a garden-hose strategy routes every function with a number of EPR pairs that is exponential in $n$~\cite{gardenhose}. A strategy built from conditional disclosure of secrets~\cite{lvw} reduces the required shared state to $2^{O(\sqrt{n\log n})}$ qubits~\cite{relating}, which is subexponential in $n$ and the best bound known for all functions~\cite{maynotes}. For lower bounds, a line of work counts only the shared state and places no limit on the messages or the local operations. For explicit functions, these bounds are linear in $n$ when recovery is perfect in one of the two routing cases~\cite{ranklower,cdsrouting}, and logarithmic when both cases may have error~\cite{robustlog}. These lower bounds leave open whether some functions need more than linearly many EPR pairs.

\subsection{Every function routes with \texorpdfstring{$2n+2$}{2n+2} EPR pairs}

This paper shows that no function needs more than linearly many EPR pairs when the messages are not limited. The price is an exponentially long quantum message from Alice. The following theorem states both facts in the model of \Cref{sec:model}.

\begin{theorem}[Linear entanglement]\label{thm:main}
Let $n\ge1$ and let $f:\{0,1\}^n\times\{0,1\}^n\to\{0,1\}$ be any Boolean function. There is a strategy for $f$-routing with the following three properties.
\begin{enumerate}
\item The shared state is $2n+2$ EPR pairs.
\item On every input pair $(x,y)$, the selected party recovers the qubit perfectly: Alice if $f(x,y)=0$, and Bob if $f(x,y)=1$.
\item Alice sends $2^{n+1}+1$ qubits and $3n+2$ classical bits, and Bob sends $3n+2$ classical bits.
\end{enumerate}
\end{theorem}

In words, $2n+2$ EPR pairs route the qubit perfectly for every Boolean function, at the price of a message from Alice that is exponentially long. The shared state is the same for all functions, and only the local operations depend on $f$. For inner product, equality, greater-than and disjointness, perfect recovery needs $\Omega(n)$ EPR pairs~\cite{ranklower,cdsrouting}. Linear upper bounds were already known for these four functions~\cite{gardenhose,ranklower}, and \Cref{thm:main} shows that no other function needs more.

\Cref{thm:main} also limits what lower bounds can show. A lower bound that counts only the shared state must hold for the strategy of \Cref{thm:main}, so it cannot exceed $2n+2$ EPR pairs. In particular, a robust lower bound, one that allows error in both routing cases, can grow at most linearly in $n$. Whether such a bound can grow faster than logarithmically remains open~\cite{maynotes,robustlog}.

\subsection{The cost moves from the shared state to the messages}

The strategy of \Cref{thm:main} trades shared entanglement for long messages. Alice's message contains two registers of $2^n$ qubits each, and she keeps two more registers of the same size. \Cref{thm:main} does not limit lower bounds that count every qubit Alice and Bob hold, including their messages. For a random function, these bounds show that linearly many qubits are needed~\cite{bcs,parallelrep}. The strategy of~\cite{relating} sends only $2^{O(\sqrt{n\log n})}$ qubits, so it remains the cheapest known for all functions when the messages are counted as well. \Cref{thm:main} thus shows that a security proof against attackers with $2n+2$ or more EPR pairs must also limit another resource, such as their communication. Whether linear entanglement can be combined with messages of subexponential length is open. For a related task with a referee, Gur, Huang and Tang~\cite{ght} use $O(n)$ pre-shared entangled qubits, but their protocol also shares random unitaries that their count leaves free, so their result does not give \Cref{thm:main}.

\subsection{The idea: a key that stays in superposition}

The strategy encrypts the unknown qubit with two key bits and lets $f(x,y)$ decide who can decrypt it, similar to the strategy of~\cite{relating}. Alice prepares an EPR pair herself. On the qubit and one half of this pair, she performs quantum teleportation~\cite{teleportation} without the measurement: she applies only its unitary part. The two bits that the measurement would produce, the key bits, therefore stay in superposition in two of her qubits. The other half of the pair, which she sends to Bob, carries the qubit encrypted with the key bits. If Bob learns both key bits, he decrypts and holds the qubit. If instead no system other than this encrypted half depends on the key bits, Alice inverts her unitary. She then holds the qubit again, and the half with Bob is one half of an unused EPR pair.

The strategy therefore has to disclose each key bit to Bob when $f(x,y)=1$ and to hide it from him when $f(x,y)=0$. The strategy of~\cite{relating} discloses the key with a classical protocol for conditional disclosure of secrets~\cite{lvw}. Such a protocol hides the key with randomness that Alice and Bob share in advance, and this randomness makes the shared state of that strategy large. The new strategy uses a quantum message instead, in which Alice hides the key with entanglement that she prepares alone. For each key bit, she prepares a maximally entangled state of two registers, each with $2^n$ qubits, one for every $n$-bit string. She sends one register to Bob and keeps the other. The strategy turns this state into the one obtained by applying $(X\pm V)/\sqrt2$ to the register sent to Bob, where the sign carries the key bit. Here $X$ is the Pauli operator that flips one of the $2^n$ qubits of this register, and $V$ is a product of Pauli $Z$ operators chosen such that $X$ and $V$ commute when $f(x,y)=1$ and anticommute when $f(x,y)=0$. If the two operators commute, the two signs give Bob orthogonal states, so he reads the key bit. If they anticommute, both signs give Bob the maximally mixed state, and Alice removes the dependence on the key bit by acting on the register she kept. So Alice is able to recover the unknown qubit.

Alice applies $X$ and $V$, but the right pair depends on both inputs. The input $x$ selects the qubits on which $V$ has a factor $Z$, and $y$ shifts this pattern relative to the qubit that $X$ flips. Alice does not have $y$ when she acts. The shared EPR pairs supply the shift. A measurement by Bob that depends on $y$ leaves Alice's halves of $n+1$ EPR pairs in a superposition of two basis states, $\ket{0,a}$ and $\ket{1,a+y}$, each labelled by a bit and an $n$-bit string, where $a+y$ is the bitwise sum modulo $2$. Alice applies $X$ to the qubit for the string $a$ on the first and $V$ shifted by $a+y$ on the second, and then measures her halves in a basis that hides which of the two she applied. Each key bit uses $n+1$ EPR pairs, which gives $2n+2$ in total.

\section{Routing model}\label{sec:model}

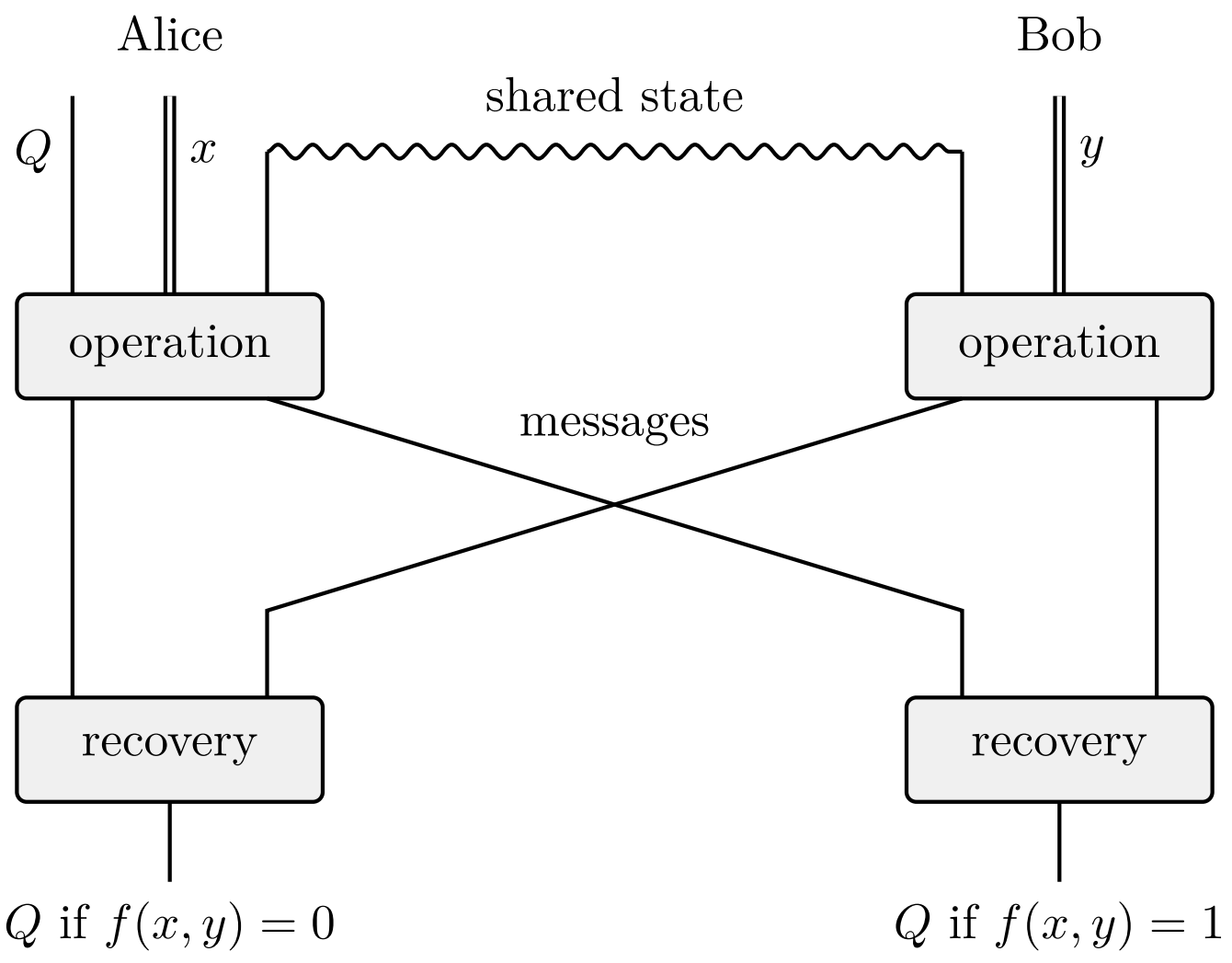
\begin{figure}[b]
\centering
\begin{tikzpicture}[thick,
 gate/.style={draw,fill=gray!12,rounded corners=2pt,minimum height=0.75cm,minimum width=2.2cm},
 lab/.style={font=\small}]
\draw[decorate,decoration={snake,amplitude=0.5mm,segment length=2.5mm}]
 (2.0,5.1) -- (7.0,5.1);
\node[lab,above] at (4.5,5.25) {shared state};
\draw (0.6,5.5) -- (0.6,4.075);
\node[lab,left] at (0.6,5.1) {$Q$};
\draw[double,double distance=1pt] (1.3,5.5) -- (1.3,4.075);
\node[lab,right] at (1.3,5.1) {$x$};
\draw (2.0,5.1) -- (2.0,4.075);
\draw[double,double distance=1pt] (7.7,5.5) -- (7.7,4.075);
\node[lab,right] at (7.7,5.1) {$y$};
\draw (7.0,5.1) -- (7.0,4.075);
\node[gate,lab] at (1.3,3.7) {operation};
\node[gate,lab] at (7.7,3.7) {operation};
\draw (0.6,3.325) -- (0.6,1.175);
\draw (8.4,3.325) -- (8.4,1.175);
\draw (2.0,3.325) -- (7.0,1.8) -- (7.0,1.175);
\draw (7.0,3.325) -- (2.0,1.8) -- (2.0,1.175);
\node[lab] at (4.5,3.1) {messages};
\node[gate,lab] at (1.3,0.8) {recovery};
\node[gate,lab] at (7.7,0.8) {recovery};
\draw (1.3,0.425) -- (1.3,-0.15) node[lab,below] {$Q$ if $f(x,y)=0$};
\draw (7.7,0.425) -- (7.7,-0.15) node[lab,below] {$Q$ if $f(x,y)=1$};
\node[lab] at (1.3,5.95) {Alice};
\node[lab] at (7.7,5.95) {Bob};
\end{tikzpicture}
\caption{A strategy for $f$-routing, with time running downward. Double lines denote classical inputs, and the wavy line denotes the shared state. Alice recovers the qubit when $f(x,y)=0$, and Bob recovers it when $f(x,y)=1$.}\label{fig:protocol}
\end{figure}

This section defines $f$-routing and perfect recovery in the model of~\cite{maynotes,complexity,robustlog}. Fix a Boolean function $f:\{0,1\}^n\times\{0,1\}^n\to\{0,1\}$. Before any input arrives, Alice and Bob share a quantum state. Alice then receives $x\in\{0,1\}^n$ and a qubit $Q$, while Bob receives $y\in\{0,1\}^n$. The shared state does not depend on $x$, $y$ or the state of $Q$. \Cref{fig:protocol} shows a strategy for $f$-routing, which consists of the shared state and the operations of Alice and Bob. Before the message exchange, Alice, who knows only $x$, applies a quantum operation to $Q$ and her part of the shared state, and Bob, who knows only $y$, applies one to his part. Each party keeps a part of the result and sends the rest to the other party. The two messages are sent simultaneously, so neither party has the other's message when it produces its own. After the exchange, no further communication takes place. If $f(x,y)=0$, Alice has to recover the qubit $Q$, and if $f(x,y)=1$, Bob has to recover the qubit $Q$. Recovery is perfect on the input pair $(x,y)$ when the selected party ends with the qubit in its original state. If the qubit was entangled with another system, the joint state of the recovered qubit and this system must be the original one as well. Only the shared state counts as a cost in this model. The operations are arbitrary, and the messages may be quantum and of any length.

\section{The strategy}\label{sec:protocol}

This section describes the strategy of \Cref{thm:main}, and \Cref{sec:proof} proves that it recovers the qubit. \Cref{subsec:notation} fixes the notation. \Cref{subsec:teleport} recalls how teleportation encrypts the unknown qubit with a key that stays in superposition. \Cref{subsec:work} defines the operators that Alice applies to her registers and shows that $f(x,y)$ decides whether they commute. \Cref{subsec:steps} lists the steps of the strategy.

\subsection{Notation}\label{subsec:notation}

Strings in $\{0,1\}^n$ are added bitwise modulo $2$, and $\ell\cdot y=\ell_1y_1+\dots+\ell_ny_n$ modulo $2$ is the inner product of two strings. We write $N=2^n$ for the number of strings. The Pauli operators are $X=\ketbra01+\ketbra10$ and $Z=\ketbra00-\ketbra11$. An EPR pair is the two-qubit state $\ket\Phi=(\ket{00}+\ket{11})/\sqrt2$. For a register $W$ of dimension $D$ with basis states $\ket w$ and a second register $W'$ of the same dimension, the maximally entangled state is
\begin{equation}\label{eq:omega}
 \ket\Omega_{WW'}=\frac1{\sqrt D}\sum_{w}\ket w_W\ket w_{W'} .
\end{equation}

\subsection{Teleportation without measurement}\label{subsec:teleport}

Alice encrypts the unknown qubit $Q$ by teleporting it without measuring. For a pair of bits $k=(k_1,k_2)$, let $P_k=X^{k_1}Z^{k_2}$. Alice prepares an EPR pair $\ket\Phi_{TC}$ of her own. The four states $\ket{\Phi_k}=(I\otimes P_k)\ket\Phi$ form the Bell basis of two qubits, and for every state $\ket\psi$ of $Q$~\cite{teleportation},
\begin{equation}\label{eq:teleport-identity}
 \ket\psi_Q\otimes\ket\Phi_{TC}
 =\frac12\sum_{k\in\{0,1\}^2}\ket{\Phi_k}_{QT}\otimes P_k\ket\psi_C .
\end{equation}
In teleportation, Alice measures $QT$ in the Bell basis, and her outcome $k$ tells the holder of $C$ which operator $P_k$ to remove. In the strategy of \Cref{thm:main}, she does not measure $QT$. She applies the \emph{Bell-basis change}, the unitary $\sum_k\ketbra{k}{\Phi_k}$ on $QT$ that maps each Bell state $\ket{\Phi_k}$ to the basis state $\ket k=\ket{k_1}\ket{k_2}$. We call the two qubits after this unitary the \emph{key qubits} $K_1$ and $K_2$, because they hold the key bits $k_1$ and $k_2$. By the teleportation identity \eqref{eq:teleport-identity}, the state is then
\begin{equation}\label{eq:teleport}
 \frac12\sum_{k\in\{0,1\}^2}\ket{k_1}_{K_1}\ket{k_2}_{K_2}\otimes P_k\ket\psi_C .
\end{equation}
The qubit $C$ carries $\ket\psi$ encrypted by $P_k$, and the key $k$ is in superposition. Two facts follow. First, a party that holds $C$ and learns $k$ recovers $\ket\psi$ by applying $P_k^\dagger$. Second, as long as the state has the form \eqref{eq:teleport}, Alice recovers $\ket\psi$ in $Q$ by inverting the Bell-basis change on $K_1K_2$, wherever $C$ is.

\subsection{Operators that commute according to \texorpdfstring{$f$}{f}}\label{subsec:work}

For each key qubit, Alice uses two \emph{work registers} $W$ and $W'$ of $N=2^n$ qubits each. She later sends $W$ to Bob, which makes her message exponentially long. The qubits of each work register are indexed by the strings $c\in\{0,1\}^n$, and we call $c$ the address of its qubit. We write $X_c$ and $Z_c$ for the Pauli operators $X$ and $Z$ on the qubit with address $c$, in $W$ or in $W'$. For Alice's input $x$ and a string $b\in\{0,1\}^n$, let
\begin{equation}\label{eq:V}
 V_{x,b}=\prod_{c\,:\,f(x,b+c)=0}Z_c .
\end{equation}
The operator $V_{x,b}$ has a factor $Z_c$ wherever the row $f(x,\cdot)$ of the function, shifted by $b$, is zero. Both $X_a$ and $V_{x,b}$ are real symmetric matrices that square to the identity. The next lemma shows that $X_a$ and $V_{x,a+y}$ anticommute when $f(x,y)=0$ and commute when $f(x,y)=1$, for every address $a$.

\begin{lemma}[Commutation]\label{lem:commute}
For all $x,y,a\in\{0,1\}^n$,
\begin{equation}
 X_aV_{x,a+y}=
 \begin{cases}
 -V_{x,a+y}X_a, & f(x,y)=0,\\
 \phantom{-}V_{x,a+y}X_a, & f(x,y)=1.
 \end{cases}
\end{equation}
\end{lemma}

\begin{proof}
The operator $X_a$ anticommutes with $Z_a$ and commutes with $Z_c$ for every $c\ne a$. By the definition \eqref{eq:V} of $V_{x,a+y}$, this product contains the factor $Z_a$ when $f(x,(a+y)+a)=0$, and not otherwise. Since $(a+y)+a=y$, the two operators anticommute when $f(x,y)=0$ and commute when $f(x,y)=1$.
\end{proof}

\subsection{The steps of the strategy}\label{subsec:steps}

The strategy treats each key qubit in its own block. Block $i\in\{1,2\}$ consists of the key qubit $K_i$, two shared registers $L_i$ and $R_i$, and the work registers $W_i$ and $W_i'$. Alice holds $L_i$ and Bob holds $R_i$, and each of these two registers has $n+1$ qubits. A basis state $\ket{r,j}$ of such a register has a label $(r,j)$ that consists of a bit $r$ and a string $j\in\{0,1\}^n$. The registers $L_i$ and $R_i$ share $n+1$ EPR pairs,
\begin{equation}\label{eq:block-state}
 \ket\Phi^{\otimes(n+1)}_{L_iR_i}
 =\frac1{\sqrt{2N}}\sum_{r\in\{0,1\}}\sum_{j\in\{0,1\}^n}\ket{r,j}_{L_i}\ket{r,j}_{R_i} .
\end{equation}
The strategy uses two orthonormal bases of a register of $n+1$ qubits. Bob measures in the basis
\begin{equation}\label{eq:bob-basis}
 \ket{\beta^y_{a,u}}=\frac{\ket{0,a}+(-1)^u\ket{1,a+y}}{\sqrt2},
 \qquad a\in\{0,1\}^n,\ u\in\{0,1\},
\end{equation}
which depends on his input $y$. Each vector is a superposition of two labels whose strings differ by $y$, and these $2N$ vectors form an orthonormal basis. Alice measures in the Hadamard basis
\begin{equation}\label{eq:hadamard-basis}
 \ket{h_{s,\ell}}=\frac1{\sqrt{2N}}\sum_{r\in\{0,1\}}\sum_{j\in\{0,1\}^n}(-1)^{sr+\ell\cdot j}\ket{r,j},
 \qquad s\in\{0,1\},\ \ell\in\{0,1\}^n .
\end{equation}
Each vector of the Hadamard basis has an overlap of absolute value $1/\sqrt{2N}$ with every basis state $\ket{r,j}$. Alice's outcome therefore reveals nothing about the label, and the measurement leaves only signs that both parties can compute after the exchange.

\begin{strategy}\label{prot:router}
The shared state is the block state \eqref{eq:block-state} for $i=1$ and $i=2$, which is $2n+2$ EPR pairs.

\smallskip\noindent\emph{Bob before the exchange.} Bob holds $y$. For each block $i$, he measures $R_i$ in the basis \eqref{eq:bob-basis} and obtains an outcome $(a_i,u_i)$. He sends $y$ and both outcomes to Alice and keeps a copy.

\smallskip\noindent\emph{Alice before the exchange.} Alice holds $x$ and the qubit $Q$.
\begin{enumerate}
\item She prepares an EPR pair $\ket\Phi_{TC}$ and applies the Bell-basis change of \Cref{subsec:teleport} to $QT$, which gives the key qubits $K_1$ and $K_2$.
\item For each block $i$, she prepares the work registers $W_iW_i'$ in the maximally entangled state $\ket\Omega$ of \eqref{eq:omega}. She then applies a unitary $G_x$ to $K_iL_iW_i$. On a key qubit that holds the bit $z$ and a register $L_i$ in the state $\ket{r,j}$, this unitary applies $X_j$ to $W_i$ if $r=0$, and $(-1)^zV_{x,j}$ if $r=1$:
\begin{equation}\label{eq:gate}
 G_x\,\ket z_{K_i}\ket{r,j}_{L_i}\ket\varphi_{W_i}=
 \begin{cases}
 \ket z\ket{0,j}\otimes X_j\ket\varphi, & r=0,\\
 \ket z\ket{1,j}\otimes(-1)^zV_{x,j}\ket\varphi, & r=1.
 \end{cases}
\end{equation}
She measures $L_i$ in the Hadamard basis \eqref{eq:hadamard-basis} and obtains an outcome $(s_i,\ell_i)$.
\item She sends the qubit $C$, the registers $W_1$ and $W_2$ of $N=2^n$ qubits each, her input $x$ and both outcomes to Bob. She keeps $K_1$, $K_2$, $W_1'$, $W_2'$ and a copy of $x$ and of her outcomes.
\end{enumerate}

\smallskip\noindent\emph{Recovery.} Both parties now know $x$, $y$ and all outcomes. For each block $i$ and each bit $z$, let
\begin{equation}\label{eq:M}
 t_i=u_i+s_i+\ell_i\cdot y \bmod 2,
 \qquad
 M_{i,z}=\frac{X_{a_i}+(-1)^{z+t_i}V_{x,a_i+y}}{\sqrt2} .
\end{equation}
\begin{itemize}
\item If $f(x,y)=0$, Alice recovers. For each block $i$, she applies $M_{i,z}$ to $W_i'$ controlled by the key qubit $K_i$, that is, the unitary $\sum_z\ketbra zz_{K_i}\otimes M_{i,z}$ on $K_iW_i'$. She then inverts the Bell-basis change on $K_1K_2$, which returns the qubit $Q$.
\item If $f(x,y)=1$, Bob recovers. For each block $i$, he measures the observable $X_{a_i}V_{x,a_i+y}$ on $W_i$, whose eigenvalues are $1$ and $-1$. He writes his outcome as $(-1)^{e_i}$ and sets $k_i=e_i+t_i$ modulo $2$. He applies $P_k^\dagger$ with $k=(k_1,k_2)$ to $C$ and outputs $C$ as the recovered qubit $Q$.
\end{itemize}
\end{strategy}

\Cref{lem:block} below shows that the steps before the exchange leave $W_iW_i'$ in the state $(M_{i,z}\otimes I)\ket\Omega$ when the key qubit $K_i$ holds the bit $z$. Up to the known bit $t_i$, the key bit $z$ is thus the sign between $X_{a_i}$ and $V_{x,a_i+y}$. Alice's recovery uses that $M_{i,z}$ is unitary when $f(x,y)=0$, and Bob's recovery uses that $X_{a_i}V_{x,a_i+y}$ is an observable when $f(x,y)=1$. \Cref{sec:proof} proves both facts and shows that Bob's bit $k_i$ equals the key bit. In the model of \Cref{sec:model}, Alice's message consists of $C$, $W_1$, $W_2$ and her classical message, and everything that Bob keeps or sends is classical.

\section{Proof of \texorpdfstring{\Cref{thm:main}}{Theorem 1}}\label{sec:proof}

This section proves that \Cref{prot:router} has the three properties of \Cref{thm:main}. \Cref{subsec:first-round} computes the state of one block after the exchange. \Cref{subsec:recoveries} uses this state to show that Alice recovers the qubit when $f(x,y)=0$ and Bob when $f(x,y)=1$, and counts the resources.

\subsection{One block after the exchange}\label{subsec:first-round}

We consider the registers $K_i$, $L_i$, $R_i$, $W_i$ and $W_i'$ of one block, for one value $z$ of the key qubit, and drop the index $i$. Each of the work registers $W$ and $W'$ has dimension $D=2^N$.

\begin{lemma}[One block]\label{lem:block}
Fix the inputs $x$ and $y$, a bit $z$, an outcome $(a,u)$ of Bob and an outcome $(s,\ell)$ of Alice. Then
\begin{equation}\label{eq:block}
 \bigl(\bra{h_{s,\ell}}_L\otimes\bra{\beta^y_{a,u}}_R\bigr)\,G_x\,
 \ket z_K\ket\Phi^{\otimes(n+1)}_{LR}\ket\Omega_{WW'}
 =\frac{(-1)^{\ell\cdot a}}{2N}\,\ket z_K\otimes(M_z\otimes I)\ket\Omega_{WW'} ,
\end{equation}
where
\begin{equation}
 M_z=\frac{X_a+(-1)^{z+t}V_{x,a+y}}{\sqrt2}
 \qquad\text{and}\qquad
 t=u+s+\ell\cdot y \bmod 2 .
\end{equation}
The vector $(M_z\otimes I)\ket\Omega$ has norm one.
\end{lemma}

The left-hand side of \eqref{eq:block} is the state of the block after both measurements, not normalized. On the right-hand side, the prefactor has absolute value $1/(2N)$ for every outcome and does not depend on $z$, and the vector after it has norm one. Every outcome is therefore equally likely for $z=0$ and $z=1$. The outcomes thus reveal nothing about the key bit, and a superposition of its two values survives. The key bit appears only in the sign between $X_a$ and $V_{x,a+y}$ in the state of the work registers.

\begin{proof}
We compute the state of the block after both measurements, which is the left-hand side of \eqref{eq:block}, in three steps: Bob's measurement, Alice's unitary $G_x$ and Alice's measurement. Bob's measurement acts on $R$ alone, so it commutes with Alice's operations and we apply it first.

First, Bob's outcome $(a,u)$ leaves $L$ in the state that $R$ was projected on. The vector $\ket{\beta^y_{a,u}}$ of Bob's basis \eqref{eq:bob-basis} has real coefficients, so
\begin{align*}
 \bigl(I_L\otimes\bra{\beta^y_{a,u}}_R\bigr)\ket\Phi^{\otimes(n+1)}_{LR}
 &=\frac1{\sqrt{2N}}\sum_{r,j}\langle\beta^y_{a,u}\vert r,j\rangle\,\ket{r,j}_L\\
 &=\frac1{\sqrt{2N}}\cdot\frac{\ket{0,a}_L+(-1)^u\ket{1,a+y}_L}{\sqrt2} .
\end{align*}
Second, by its definition \eqref{eq:gate}, the unitary $G_x$ applies $X_a$ to $W$ in the first term and $(-1)^zV_{x,a+y}$ in the second term. This gives
\begin{equation*}
 \frac1{\sqrt{2N}}\,\ket z_K\otimes
 \frac{\ket{0,a}_L\otimes(X_a\otimes I)\ket\Omega
 +(-1)^{u+z}\ket{1,a+y}_L\otimes(V_{x,a+y}\otimes I)\ket\Omega}{\sqrt2} .
\end{equation*}
Third, Alice's outcome $(s,\ell)$ projects $L$ on the vector $\ket{h_{s,\ell}}$ of the Hadamard basis \eqref{eq:hadamard-basis}. Its overlaps with the two labels are
\begin{equation*}
 \langle h_{s,\ell}\vert0,a\rangle=\frac{(-1)^{\ell\cdot a}}{\sqrt{2N}},
 \qquad
 \langle h_{s,\ell}\vert1,a+y\rangle=\frac{(-1)^{s+\ell\cdot a+\ell\cdot y}}{\sqrt{2N}} .
\end{equation*}
The projection therefore gives
\begin{equation*}
 \frac{(-1)^{\ell\cdot a}}{2N}\,\ket z_K\otimes
 \frac{\bigl(X_a+(-1)^{z+u+s+\ell\cdot y}V_{x,a+y}\bigr)\otimes I}{\sqrt2}\,\ket\Omega .
\end{equation*}
Since $t=u+s+\ell\cdot y$ modulo $2$, the operator in front of $\ket\Omega$ is $M_z\otimes I$, so this expression is the right-hand side of \eqref{eq:block}.

It remains to compute the norm. Write $V=V_{x,a+y}$ and $\sigma=(-1)^{z+t}$. The matrix $M_z$ is real, so the squared norm of $(M_z\otimes I)\ket\Omega$ is $\Tr(M_z^{\T}M_z)/D$. Since $X_a$ and $V$ are symmetric and square to the identity,
\begin{align*}
 M_z^{\T}M_z
 &=\tfrac12\,(X_a+\sigma V)(X_a+\sigma V)\\
 &=\tfrac12\,\bigl(X_a^2+V^2+\sigma(X_aV+VX_a)\bigr)\\
 &=I+\tfrac{\sigma}{2}\,(X_aV+VX_a) .
\end{align*}
The matrix $V$ is diagonal and $X_a$ maps every basis state to a different one, so $X_aV$ and $VX_a$ have zero diagonal. Hence $\Tr(M_z^{\T}M_z)=D$, and the norm is one.
\end{proof}

\subsection{The two recoveries}\label{subsec:recoveries}

\begin{proof}[Proof of \Cref{thm:main}]
We first take the qubit $Q$ in a pure state $\ket\psi$. We combine the two blocks, and then show that Alice recovers $\ket\psi$ when $f(x,y)=0$ and Bob when $f(x,y)=1$. The part on perfect recovery extends both routing cases to a qubit that is entangled with another system, and the last part counts the EPR pairs and the messages.

\smallskip\noindent\emph{Both blocks together.} After step~(1), the key qubits and $C$ are in the encrypted state \eqref{eq:teleport}. All later operations are linear, and those of block $i$ act only on $K_i$, $L_i$, $R_i$, $W_i$ and $W_i'$. We therefore apply \Cref{lem:block} to block $i$ with $z=k_i$, in each term $k$ of \eqref{eq:teleport}. Every term gets the same prefactor, of absolute value $1/(2N)^2$. For fixed outcomes of the four measurements, the normalized state after the exchange is therefore, up to a sign,
\begin{equation}\label{eq:after-round}
 \frac12\sum_{k\in\{0,1\}^2}\ket{k_1}_{K_1}\ket{k_2}_{K_2}\otimes P_k\ket\psi_C
 \otimes(M_{1,k_1}\otimes I)\ket\Omega_{W_1W_1'}\otimes(M_{2,k_2}\otimes I)\ket\Omega_{W_2W_2'} ,
\end{equation}
with the matrices $M_{i,z}$ of \eqref{eq:M}. The vector \eqref{eq:after-round} has norm one because every vector $(M_{i,z}\otimes I)\ket\Omega$ has norm one by \Cref{lem:block}. It differs from the encrypted state \eqref{eq:teleport} only in the work registers, which now depend on the key.

\smallskip\noindent\emph{Alice recovers when $f(x,y)=0$.} We show that Alice's operations return each pair of work registers to $\ket\Omega$, which removes the dependence on the key. The proof of \Cref{lem:block} showed that $M_{i,z}^{\T}M_{i,z}=I+\tfrac\sigma2(X_{a_i}V+VX_{a_i})$, with $V=V_{x,a_i+y}$ and $\sigma=(-1)^{z+t_i}$. The second term vanishes because the two operators anticommute by \Cref{lem:commute}, so $M_{i,z}^{\T}M_{i,z}=I$. The matrix $M_{i,z}$ is real and symmetric, so it is unitary and $M_{i,z}^2=I$. Alice applies $M_{i,z}$ to $W_i'$. An operator on one half of the maximally entangled state can be moved to the other half as its transpose, $(I\otimes M)\ket\Omega=(M^{\T}\otimes I)\ket\Omega$ for every matrix $M$ written in the basis of \eqref{eq:omega}. Hence
\begin{align*}
 (I\otimes M_{i,z})(M_{i,z}\otimes I)\ket\Omega
 &=(M_{i,z}\otimes I)(I\otimes M_{i,z})\ket\Omega\\
 &=(M_{i,z}\otimes I)(M_{i,z}^{\T}\otimes I)\ket\Omega\\
 &=(M_{i,z}^2\otimes I)\ket\Omega\\
 &=\ket\Omega .
\end{align*}
Alice applies this operation controlled by the key qubit $K_i$, so in each term of \eqref{eq:after-round} it acts with $z=k_i$. The state becomes
\begin{equation*}
 \Bigl(\frac12\sum_{k\in\{0,1\}^2}\ket{k_1}_{K_1}\ket{k_2}_{K_2}\otimes P_k\ket\psi_C\Bigr)
 \otimes\ket\Omega_{W_1W_1'}\otimes\ket\Omega_{W_2W_2'} .
\end{equation*}
The first factor is the encrypted state \eqref{eq:teleport}. Inverting the Bell-basis change on $K_1K_2$ turns it into $\ket\psi_Q\otimes\ket\Phi_{TC}$ by the teleportation identity \eqref{eq:teleport-identity}, so Alice holds $\ket\psi$ in $Q$.

\smallskip\noindent\emph{Bob recovers when $f(x,y)=1$.} We show that Bob's measurement finds the key bit $k_i$ with certainty in every term of \eqref{eq:after-round}. By \Cref{lem:commute}, the operators $X_{a_i}$ and $V_{x,a_i+y}$ commute. Their product $S_i=X_{a_i}V_{x,a_i+y}$ is then a real symmetric matrix with $S_i^2=I$, so it is an observable with eigenvalues $1$ and $-1$. Write $V=V_{x,a_i+y}$ and $\sigma=(-1)^{z+t_i}$. Then
\begin{align*}
 S_i\,(X_{a_i}+\sigma V)
 &=X_{a_i}VX_{a_i}+\sigma X_{a_i}VV\\
 &=VX_{a_i}X_{a_i}+\sigma X_{a_i}\\
 &=V+\sigma X_{a_i}\\
 &=\sigma\,(X_{a_i}+\sigma V) ,
\end{align*}
where the second line uses that $X_{a_i}$ and $V$ commute and that $V^2=I$, and the last line uses $\sigma^2=1$. Hence $S_iM_{i,z}=(-1)^{z+t_i}M_{i,z}$, and $(M_{i,z}\otimes I)\ket\Omega$ is an eigenvector of $S_i$ on $W_i$ with eigenvalue $(-1)^{z+t_i}$. In the term $k$ of \eqref{eq:after-round}, Bob's measurement of $S_i$ therefore gives the outcome $(-1)^{k_i+t_i}$ with certainty and leaves the term unchanged. The four terms give four different pairs of outcomes, so the two measurements select one term $k$, each with probability $1/4$, and the bits that Bob computes from his outcomes are the key bits of this term. His correction $P_k^\dagger$ maps $P_k\ket\psi$ to $\ket\psi$, and the state becomes
\begin{equation*}
 \ket\psi_C\otimes\ket{k_1}_{K_1}\ket{k_2}_{K_2}
 \otimes(M_{1,k_1}\otimes I)\ket\Omega_{W_1W_1'}\otimes(M_{2,k_2}\otimes I)\ket\Omega_{W_2W_2'} ,
\end{equation*}
so Bob holds $\ket\psi$ in $C$.

\smallskip\noindent\emph{Perfect recovery.} In both routing cases and for all outcomes, the final state is $\ket\psi$ in the recovered qubit, in tensor product with a state of the other registers that does not depend on $\ket\psi$. Every operation of the strategy is linear and does not depend on $\ket\psi$ either. The same calculation therefore applies to a qubit $Q$ that is entangled with another system, and the strategy returns the joint state of this system and the qubit unchanged. Hence recovery is perfect on every input pair $(x,y)$, which is the second property of \Cref{thm:main}.

\smallskip\noindent\emph{Resources.} The shared state is the block state \eqref{eq:block-state} for two blocks, which is $2(n+1)=2n+2$ EPR pairs. The EPR pair $TC$ and the work registers are prepared by Alice alone and are not part of the shared state. Alice sends the qubit $C$ and the registers $W_1$ and $W_2$ of $N=2^n$ qubits each, in total $2^{n+1}+1$ qubits. She also sends $x$ and two outcomes $(s_i,\ell_i)$, which are $n+2(n+1)=3n+2$ bits. Bob sends $y$ and two outcomes $(a_i,u_i)$, which are $3n+2$ bits.
\end{proof}

\section{Conclusion}\label{sec:conclusion}

This paper showed that $2n+2$ shared EPR pairs suffice to route a qubit perfectly for every Boolean function, so a lower bound that counts only the shared state can grow at most linearly in $n$. The price is an exponentially long quantum message from Alice. Two problems remain open. First, is there a function for which routing with constant error in both routing cases needs $\Omega(n)$ EPR pairs when the messages are not limited? By \Cref{thm:main}, such a bound would be optimal. Second, can the messages be shorter? The strategy sends $2^{n+1}+1$ qubits, and it is open whether every function has a strategy with $O(n)$ EPR pairs and messages of subexponential length.

\section*{Lean formalization}

A Lean formalization proves \Cref{thm:main}, using only Lean's three standard axioms. The code is available at \url{https://github.com/kevinbogner/linear-routing-lean}.

\section*{Disclosure of AI use}

The strategy was found by GPT-6.2 Sol (OpenAI) in a research session directed by the author, and the same model wrote the Lean formalization. Claude Fable 5.2 (Anthropic) checked the strategy by independent derivation and numerical simulation, drafted the text and reviewed the formal statement. The author takes responsibility for the paper's content.

\providecommand{\etalchar}[1]{$^{#1}$}

\end{document}